\documentclass[11pt,a4paper]{article}
\usepackage{fullpage}

\usepackage{latexsym}

\usepackage{color}
\usepackage{graphicx}
\usepackage{listings}
\usepackage{url}

\usepackage{tikz}
\usetikzlibrary{shadows,arrows,shapes,positioning,calc}
\tikzset{
ell/.style={draw,ellipse,minimum height=2em,minimum width=8em,align=center},
line/.style={-,line width=0.5pt},
dashline/.style={-,dashed,line width=0.5pt},
dot/.style = {draw,fill,circle,inner sep=0pt,outer sep=0pt,minimum size=2pt},
}

\newcommand{\floor}[1]{\lfloor #1\rfloor}

\newtheorem{proposition}{Proposition}
\newenvironment{proof}{\emph{Proof:}}{$\Box$\newline}

\title{The Shortest Path Problem with Edge Information Reuse is NP-Complete}
\date{15.9.2015, Revised 6.6.2016}
\author{Jesper Larsson Tr\"aff\\
TU Wien\\
Faculty of Informatics, Institute of Information Systems\\ 
Research Group Parallel Computing\\
Favoritenstrasse 16/184-5, 1040 Vienna, Austria\\
email: \texttt{traff@par.tuwien.ac.at}
}

\begin{document}
\maketitle

\begin{abstract}
We show that the following variation of the single-source shortest
path problem is NP-complete. Let a weighted, directed, acyclic graph
$G=(V,E,w)$ with source and sink vertices $s$ and $t$ be given. Let in
addition a mapping $f$ on $E$ be given that associates information with
the edges (e.g., a pointer), such that $f(e)=f(e')$ means that edges
$e$ and $e'$ carry the same information; for such edges it is required
that $w(e)=w(e')$. The length of a simple $st$ path $U$ is the sum of
the weights of the edges on $U$ but edges with $f(e)=f(e')$ are counted
only once. The problem is to determine a shortest such $st$ path. We
call this problem the \emph{edge information reuse shortest path problem}. It
is NP-complete by reduction from 3SAT.
\end{abstract}

\section{The Edge Information Reuse Shortest Path Problem}

A weighted, directed, acyclic graph $G=(V,E,w)$ with source and sink
vertices $s,t\in V$ is given. Edges represent some possible
substructures and a path from $s$ to $t$ determines how substructures
are put together to form a desired superstructure. Edge weights
reflect the cost of the associated substructures (e.g., memory
consumption). An ordinary shortest path determines an ordered,
tree-like representation of the superstructure of least cost: a single
root with children given by the edges of the path. Now, different
edges in $G$ may represent similar substructures, and if several such
edges occur on a path, the cost of the substructure need be paid only
once. We want to find a shortest such path from $s$ to $t$ where the
weight of edges with similar information is counted only once.  There
may easily be such a path of less cost than an ordinary shortest path
in $G$. In that case, a more cost-efficient representation of the
desired superstructure is by a compressed tree where several children
are represented by the same substructure. We model this problem by
associating information with the edges in the form of a function $f$
on $E$ that may for instance be a pointer to the substructure
represented by the edge. Two edges $e,e'\in E$ carry the same
information and are similar in that respect if $f(e)=f(e')$. For the
optimization problem to be well-defined, it is required that edges
with $f(e)=f(e')$ have the same weight $w(e)=w(e')$ (alternatively, 
the edge with the smallest weight could be counted). Weights and
pointers are positive integers.

We call this problem the \emph{edge information reuse shortest path
  problem}. The decision version is formalized as follows.

\medskip
\noindent
INSTANCE: Given a weighted, directed, acyclic graph $G=(V,E,w)$, two
vertices $s,t\in V$, edge information function $f$, integer $K\geq 0$.
The length $r(U)$ of a(n ordered) path $U$ with edge information reuse
is defined as
\begin{eqnarray*}
r(U) & = & \sum_{e\in U\setminus \{e\in U | \exists e'<e: f(e')=f(e)\}}w(e)
\end{eqnarray*}
where $e'<e$ means that edge $e'$ occurs before $e$ on $U$.

\medskip
\noindent
QUESTION: Is there an $st$ path $U$ in $G$ with length $r(U)\leq K$?

\medskip
The edge information reuse shortest path problem is NP-complete, as the next
section will show.

\section{Reduction from 3SAT}

Originally, a reduction from PARTITION was given, which turned out to
be incorrect and fundamentally flawed (see Appendix). Instead, we use
the same reduction from 3SAT~\cite[LO2]{GareyJohnson79} as in the
paper by Broersma et al.~\cite{BroersmaLiWoegingerZhang05}.

\begin{proposition}
The edge information reuse shortest path problem is NP-complete.
\end{proposition}

\begin{proof}
It is clear that the problem is in NP. Let $U$ be a path from $s$ to
$t$. We can check whether the path taking edge information reuse into
account has length at most some $K$ by two traversals of $U$. We use
$f(e)$ as a pointer to mark for information reuse. In the first
traversal, all $f(e)$ for $e\in U$ are marked as unused. In the second
traversal the weights of unused $f(e)$'s are summed and $f(e)$ marked
as used.

The 3SAT problem is, given a collection of $m$ three-literals clauses
$C_j, 0\leq j<m$ over $n$ variables $x_i,0\leq i<n$, to determine
whether there is a true/false assignment to the $n$ variables that
satisfies the instance by making all $m$ clauses true.

Let an instance of 3SAT be given. For each variable $x_i$ and each
clause $C_j$ a weighted subgraph is constructed, and these subgraphs
are joined one after the other. All edge weights will be either 0 or 1
(not actually corresponding to truth values). The first node of the
subgraph corresponding to $x_0$ is taken as the source $s$, and the
last node of the subgraph corresponding to $C_{m-1}$ is taken as the
sink $t$. Any $st$ path will have to pass through all variable and all
clause subgraphs. The part of the path passing through the variable
subgraphs will correspond to an assignment of truth values to the
variables $x_i$. Edges in the variable subgraphs corresponding to
$x_i$ will share information with edges in the clause subgraphs where
a literal corresponding to $x_i$ appears. The construction will have
the property that there is an information reuse shortest path of
length $n$ if and only if the 3SAT instance is satisfiable.

The subgraph for variable $x_i$ consists of four vertices
$u_i,u_i',\bar{u}'_i, u_{i+1}$ for $0\leq i<n$. The source vertex is
$s=u_0$. There are edges $(u_i,u'_i)$ and $(u_i,\bar{u}'_i)$ of weight
1, and edges $(u'_i,u_{i+1})$ and $(\bar{u}'_i,u_{i+1})$ of weight
0. A path going through vertex $u'_i$ will assign true to $x_i$,
whereas a path going through vertex $\bar{u}'_i$ will assign true to
the negation $\bar{x}_i$.  The edges $(u_i,u'_i)$ and
$(u_i,\bar{u}'_i)$ will share information with edges in clause
subgraphs.

The subgraph for clause $C_j$ consists of five vertices
$v_j,v^0_j,v^1_j,v^2_j,v_{j+1}$ for $0\leq j<m$. The sink vertex is
the last vertex $t=v_m$. The last vertex $u_n$ of the variable
subgraph is connected to the first vertex $v_0$ of the clause subgraph
by an edge $(u_n,v_0)$ of weight 0. In each clause subgraph, there are
three edges of weight 0, namely $(v^0_j,v_{j+1})$, $(v^1_j,v_{j+1})$,
$(v^2_j,v_{j+1})$. Likewise there are three edges of weight 1, namely
$(v_j,v^0_j)$, $(v_j,v^1_j)$, $(v_j,v^2_j)$ which share information
with edges from the variable subgraphs.  Information sharing is as
follows: $f((u_i,u'_i)) = f((v_j,v^k_j))$ if literal $k,0\leq k<3$ of
clause $C_j$ is $x_i$, and $f((u_i,\bar{u}'_i)) = f((v_j,v^k_j))$ if
literal $k,0\leq k<3$ of clause $C_j$ is $\bar{x}_i$.

The graph thus constructed has $O(n+m)$ vertices and edges, and can
clearly be computed in a polynomial number of steps, including the
function $f$ assigning (shared) information with the edges.

Now, the claim is that the given 3SAT instance is satisfiable, if an
only if the shortest $st$ path with edge information reuse has length
exactly $n$. Any path from $s$ to $t$ must have length at least $n$,
since it has to pass through all vertices $u_i, 1\leq i<n+1$ via an
edge with weight 1. If the 3SAT instance is satisfiable, a path of
length $n$ where the weights of edges that share information with
previous edges are not counted can be constructed as follows. The path
through the variable subgraph is given directly by the truth
assignment (if $x_i$ is true, pass through $u'_i$, if $\bar{x}_i$ is
true, pass through $\bar{u}'_i$). For each clause subgraph, at least
one literal must be true, and choosing the corresponding edge for the
path costs nothing on the path length since this edge can share
information with the edge of the literal.  Conversely, if there is an
edge information reuse shortest path of length $n$, this path shows
that a truth assignment exists that satisfies all clauses since it must
pass through the clause subgraph through edges that share information with
edges (set to true) in the variable subgraph.
\end{proof}

The graph constructed in the reduction is acyclic, which shows that
the shortest $st$ path problem with edge information reuse on acyclic
graphs is NP-complete.

\section{Remark}

Alex Khodaverdian (personal communication, 24.5.2016) pointed out some
problems with the original reduction from PARTITION (still retained in
the appendix), which eventually showed this reduction to be
fundamentally flawed. Fortunately, the 3SAT reduction
from~\cite{BroersmaLiWoegingerZhang05} shows the problem NP-complete,
so that the original claim still holds. It might still be enlightening
to find a different reduction.

For the first version of this note, Gerhard Woeginger (personal
communication, 15.9.2015) pointed out that NP-completeness follows
straightforwardly from the NP-completeness of the problem of finding a
simple $st$ path with the smallest number of colors in a graph with
colored edges, and supplied the
reference~\cite{BroersmaLiWoegingerZhang05}. Simply let the function
$f$ denote the edge colors and take all edge weights to be $1$. A
shortest $st$ path with edge information reuse corresponds to a simple
$st$ path with the smallest number of colors.

\bibliographystyle{plain}
\bibliography{traff,parallel}

\appendix
\section{The flawed reduction from PARTITION}

We tried to show the edge information reuse shortest path problem
NP-complete by a reduction from the NP-complete PARTITION
problem~\cite[SP12]{GareyJohnson79}.

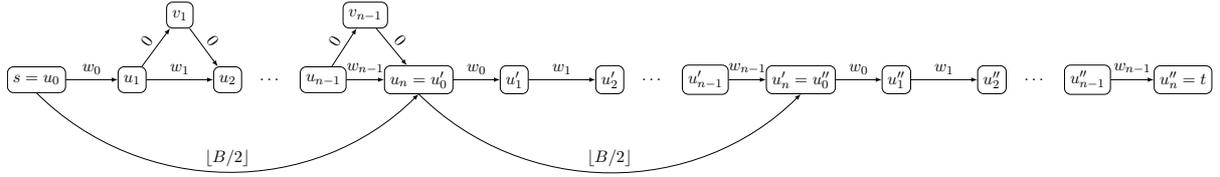
\begin{figure}
\resizebox{\textwidth}{!}{%
\begin{tikzpicture}
	\def \udist {2.2cm}
	\def \vshift {1.5cm}
	\def \graphnodeheight {.6cm}

	\tikzset{
		graphnode/.style = {
			draw, rounded corners, align=center,
			minimum height=\graphnodeheight
		},
		edgenode/.style = {midway, anchor=south, sloped},
		arrowedge/.style = {->,>=latex}
	}

	\node[graphnode] (u0) at (0,0) {$s=u_0$};
	\def \prevnode {u0}
	\foreach \name/\label/\scaledist in {u1/u_1/1, u2/u_2/1, un-1/u_{n-1}/1,
		un/u_n=u_0'/1, u1'/u_1'/1, u2'/u_2'/1, un-1'/u_{n-1}'/1,
		un'/u_n'=u_0''/1, u1''/u_1''/1, u2''/u_2''/1, un-1''/u_{n-1}''/1,
		un''/u_n''=t/1}
	{
		\path let \p1=(\prevnode) in
			node[graphnode] (\name) at
				({\x1 + \scaledist * \udist},0) {$\label$};

		\global\let\prevnode\name
	}

	\foreach \lnode/\rnode/\label in {u0/u1/w_0, u1/u2/w_1, un-1/un/w_{n-1},
		un/u1'/w_0, u1'/u2'/w_1, un-1'/un'/w_{n-1}, un'/u1''/w_0,
		u1''/u2''/w_1, un-1''/un''/w_{n-1}}
	{
		\draw[arrowedge] (\lnode) -- node[edgenode] {$\label$} (\rnode);
	}

	\foreach \lnode/\rnode in {u2/un-1, u2'/un-1', u2''/un-1''}
	{
		\path (\lnode) -- node[midway] {\dots} (\rnode);
	}

	\foreach \lnode/\rnode/\name/\label in {u1/u2/v1/v_1,
		un-1/un/vn-1/v_{n-1}}
	{
		\path (\lnode) -- node[graphnode,midway,yshift=\vshift] (\name)
			{$\label$} (\rnode);
		\draw[arrowedge] (\lnode) -- node[edgenode] {$0$} (\name);
		\draw[arrowedge] (\name) -- node[edgenode] {$0$} (\rnode);
	}

	\foreach \lnode/\rnode in {u0/un, un/un'}
	{
		\draw (\lnode.south) edge[arrowedge,out=315,in=225] node[edgenode]
			{$\floor{B/2}$} (\rnode.south);
	}
\end{tikzpicture}
}
\caption{The weighted, directed, acyclic graph constructed from the
  given PARTITION instance. The shortest path taking edge information
  reuse into account should have length $B$ if and only if the PARTITION
  instance is feasible.}
\label{fig:graphconstruction}
\end{figure}

\begin{proposition}
The edge information reuse shortest path problem is NP-complete.
\end{proposition}

It is clear that the problem is in NP. Let $U$ be a path from $s$ to
$t$. We can check whether the path taking edge information reuse into
account has length at most some $K$ by two traversals of $U$. We use
$f(e)$ as a pointer to mark for information reuse. In the first
traversal, all $f(e)$ for $e\in U$ are marked as unused. In the second
traversal the weights of unused $f(e)$'s are summed and $f(e)$ marked
as used.

Now, let the item set $A=\{x_0,\ldots x_{n-1}\}$ with integer item
weights $w(x_i)$ be an instance of PARTITION, and let $B=\sum_{x \in
  A} w(x)$.  We construct a weighted, directed, acyclic graph $G$ such
that there is a shortest path with edge information reuse of length
$B$ if and only if the answer to PARTITION is ``yes'', that is, if
there is a subset $P$ of $A$ such that $\sum_{x\in P}w(x)=\sum_{x\in
  A-P}w(x)$.

The graph $G$ consists if three parts, the first with vertices
$u_0,u_1,\ldots, u_n$ and $v_1,\ldots v_{n-1}$, the second with
vertices $u'_0,u'_1,\ldots, u'_n$, and the third part with vertices
$u''_0,u''_1,\ldots, u''_n$, where $u_n=u'_0$ and $u'_n=u''_0$. The
source is $s=u_0$ and the sink $t=u''_n$. For each $i=0,n-1$ there are
edges $(u_i,u_{i+1})$, $(u'_i,u'_{i+1})$, and $(u''_i,u''_{i+1})$ each
with weight $w(x_i)$, and these three edges carry the same item
information, that is
$f((u_i,u_{i+1}))=f((u'_i,u'_{i+1}))=f((u''_i,u''_{i+1}))$. Different
edges $(u_i,u_{i+1})$ and $(u_j,u_{j+1})$ carry different item
information. The information associated with these edges designate the
items of $A$. There are edges $(u_0,u_n)$ and $(u'_0,u'_n)$ both of
weight $\floor{B/2}$ (which share no information). In the first part
of $G$ there is for each $i=1,\ldots,n-1$ two edges $(u_i,v_i)$ and
$(v_i,u_{i+1})$ of weight $0$. The construction is illustrated in
Figure~\ref{fig:graphconstruction}. It can clearly be carried out in
$O(n)$ operations.

There is clearly a path from $s$ to $t$ in $G$. The purpose of first
part of the graph is to select items for the set $P$. For any path
through this part the information associated with non-zero weight
edges designate items chosen for $P$. Each subset $P$ is chosen by
a unique path.

Assume that a shortest $st$ path with information reuse is longer than
$B$. The path either passes through the vertices
$u_1,u_2,\ldots,u_{n-1}$, and possibly some of the $v_i$ vertices as
well, or takes the shortcut edge $(u_0,u_n)$.

In the first case, the length of the path to vertex $u_n$ is at most
$\floor{B/2}$, and there is no possibility for information reuse since
all edge information occurs for the first time in this part of
$G$. Let $W$ be the sum of the weights of the $(u_i,u_{i+1})$ edges on
the path. These weights will not be counted again since they share
information with the edges $(u'_i,u'_{i+1})$ and $(u''_i,u''_{i+1})$.
If $W<\floor{B/2}$ the shortest path must take the shortcut edge
$(u'_0,u'_n)$ of weight $\floor{B/2}$ since $B-W>\floor{B/2}$. The
path through the final part of $G$ from $u''_0$ to $t$ will have
weight $W-B$ for a total of $B+\floor{B/2}$.

If the shortest path instead takes the shortcut edge $(u_0,u_n)$ of
weight $\floor{B/2}$ (which means that there is no item $x_i\in A$
with $w(x_i)\leq\floor{B/2}$), then the path must also take the
shortcut edge $(u'_0,u'_n)$ since there is no edge information to
reuse and the non-shortcut path from $u'_0$ to $u'_n$ has length
$B$. The part of the path from $u''_0$ to $t$ in this case has cost
$B$ since no edge information can be reused, for a total of
$2\floor{B/2}+B$.

Thus, the length of a shortest path with information reuse is $B$ if
and only it is possible to find a path from $u_0$ to $u_n$ of length
exactly $\floor{B/2}$ (that does not use the shortcut edge), and if
the weight of the edges whose information has not been used in this
part is likewise exactly $\floor{B/2}$. In this case, all edge
information has been used, and the part of the path from $u''_0$ to
$t$ contributes weight $0$. The solution $P$ to the PARTITION is the
information $f(e)$ on the non-zero edges on the part of the path from
$s$ to $u_n$.

\medskip

The flaw in this argument is simply that subpaths of a shortest
$st$-path do not have to be shortest paths themselves. Therefore, with
information reuse, there is always a shortest path of length exactly
$B$, and the shortcut edges of weight $\floor{B/2}$ do not have to be
taken.

\end{document}